\documentclass[aps,prb,preprint,groupedaddress]{revtex4}
\usepackage{graphicx}
\usepackage{amsmath}
\usepackage{amssymb}
\usepackage{color}

\begin{document}

\title{A Search for Defect Related Ferromagnetism in SrTiO$_3$}

\author{D.A. Crandles,  B. DesRoches, F. S. Razavi}

\affiliation{Department of Physics, Brock University, St.
Catharines, Ontario, Canada, L2S 3A1} \email{dcrandles@brocku.ca}

\date{\today}

\begin{abstract}
Room temperature ferromagnetic hysteresis is observed in commercial SrTiO$_3$ substrates purchased from a variety of suppliers. It is shown that the ferromagnetic signal comes from the unpolished surfaces. Surface impurity phases cannot be detected using either x-ray diffraction or energy dispersive x-ray spectra on the unpolished surfaces. However, a possible correlation between surface disorder (xray diffraction peak linewidth) and ferromagnetism  is observed.  Ar ion bombardment (10keV-90 keV) can be used to produce surface layer disorder but is not found to induce ferromagnetism.  Annealing of the substrates at temperatures ranging from 600 to 1100 $^\circ$C is found to alter the hysteresis curves differently depending on whether the annealing is performed in air or in vacuum.  Identical annealing behaviour is observed if the substrates are artificially spiked with iron.   This suggests that  the ferromagnetic hysteresis of as purchased SrTiO$_3$ could be due to Fe contamination of the unpolished surfaces. In addition, it is shown that no ferromagnetism is observed in samples that contain 10$^{19}$-10$^{21}$ cm$^{-3}$ of oxygen vacancies if all the faces are polished.

\end{abstract}
 \maketitle

\section{Introduction}
Claims of room temperature ferromagnetism (FM)  have been made for  several  semiconducting oxides that do not contain any magnetic ions.     The phenomenon has been observed in a variety of binary oxides prepared in nanoparticle\cite{sundaresan}  and thin-film form \cite{yoon_jcm,yoon,hong_tio2,rumaiz,sudakar, coey_prb,hong_zno}.    There has been considerable interest in these otherwise non-magnetic oxides as a part of the effort to develop suitable materials for spintronic devices.  

This work is concerned with the origin of room temperature magnetic hysteres is observed in single crystal SrTiO$_3$ (001) substrates (STO)  purchased from a variety of suppliers.   The  hysteresis, which cannot be explained by non-interacting paramagnetic impurities, is interesting in light of recent work which highlights the considerable potential of STO surfaces for electronic and spintronic devices.   Consider, for example,   the two dimensional electron gas formed at the interfaces of  LaAlO$_3$/STO heterostructures \cite{ohtomo} as well as other phenomena such as blue luminescence \cite{kan} and high conductivity \cite{reagor}  produced by Ar$^+$ irradiation.  

The observation of FM hysteresis in commercial STO raises several questions, which include whether the hysteresis is a bulk or surface effect, whether or not it can be associated with an impurity phase, or whether or not it can be associated with a particular kind of point or extended defect that can be controlled through materials processing. It has been established that the surfaces of STO can be quite complex.  For example, slicing a Verneuil grown STO crystal into substrates suitable for thin film growth produces a large concentration of dislocations and voids \cite{shapiro} near the cut surface while annealing in oxidizing or reducing conditions can promote  segregation into Ti-rich or Sr-rich regions in the surface skin layer\cite{szot}.      In the experiments described below, particular attention is paid to the effect of annealing in reducing and oxidizing atmospheres  since it has been found in other oxides where FM has been observed, for example TiO$_2$\cite{sudakar} and HfO$_2$\cite{coey_prb}, that the magnitude of the remanent moment can be altered by vacuum annealing.  

It is interesting to note that in STO oxygen vacancies donate free electrons to the hybridized Ti(3d)-O(2p) conduction band and can produce superconductivity with maximum T$_c \approx$  0.3 K  \cite{koonce}.  The possibility of the coexistence of ferromagnetism and superconductivity is a compelling reason to study the magnetic properties of reduced STO.   Finally,  STO is a common substrate for metal oxide thin film growth. Since  certain thin film materials with interesting magnetic and charge ordering properties such as LaVO$_3$ \cite{fujioka} can only be produced in reducing atmospheres \cite{hotta}, it is important to be aware of how reduction at elevated temperatures affects the magnetic properties of the substrate.

\section{Results and Discussion}

\subsection{Bulk or Surface Effect?}
 
STO (100) single crystal substrates were purchased from a variety of suppliers:  Alfa Aesar,  Crystec ,  MTI corporation and  Semiconductor Wafer Inc.  Substrates were purchased in both one side polished (1sp) and two side polished (2sp) form.  Extreme care was taken - for example always using teflon tweezers that had never been in contact with stainless steel - to avoid contamination of the samples with magnetic elements\cite{abraham}.  Magnetic moment measurements were performed using a Quantum Design SQUID magnetometer.   Some small degree of magnetic hysteresis at 300 K was  measured in every single one of the over 50 substrates tested.

\begin{figure}
 \includegraphics{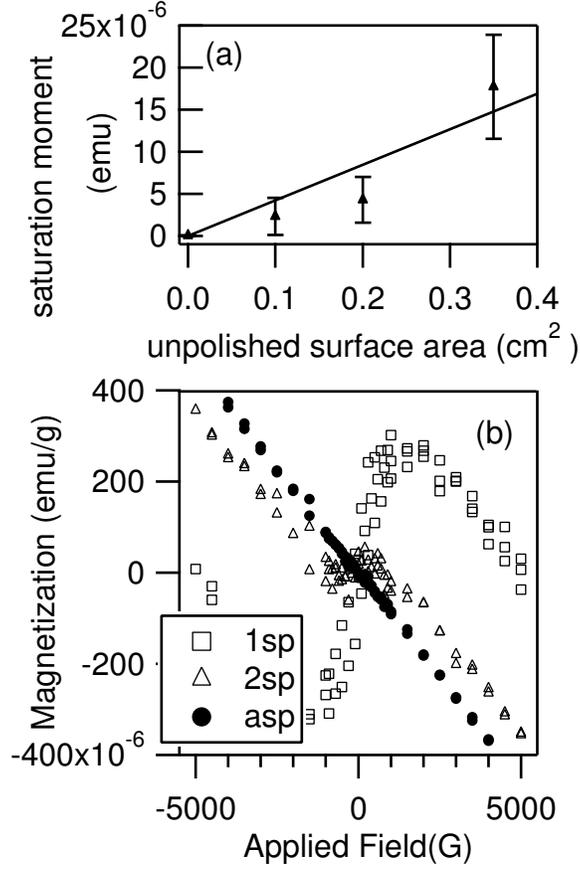}
\caption{(a) The average saturation moment as a function of unpolished surface area. The uncertainty bars represent the sample to sample standard deviation.   For example, for the point where the unpolished surface area is 0.35 cm$^2$, which is for  1sp substrates 0.5 mm thick, the data are for 19 different samples from two different suppliers.  For the point where the unpolished surface area is approximately 0.2 cm$^2$ which is for  2sp substrates 1 mm thick, the data are for 15 different samples from four different suppliers.  (b) Magnetization versus applied field at 300 K for a 5$\times$5 mm  STO  substrate polished on one side (open squares marked 1sp), hand polished on  both 5$\times$5 mm faces ( open triangles marked 2sp), polished on all six faces of the parallelpiped ( closed circles marked asp).  }
\label{polishing}
\end{figure}

To establish whether the hysteresis was a bulk or a surface effect, two sorts of tests were done.  HCl and HNO$_3$ acids  were both found to reduce the size of the remanent moment while polishing the unpolished surfaces in stages down to mirror-like smoothness was found to completely remove all traces of FM hysteresis.  Fig. \ref{polishing}(b)  compares measurements of moment versus field  on  STO substrates  with one side polished (1sp) - the as purchased state - with substrates with both $5\times5$ mm surfaces polished (2sp) as well as substrates with all six sides of the substrate polished (asp).  Each round of polishing removed a surface layer approximately 10 to 15 $\mu m$ thick.  The hysteresis is clearly associated with the unpolished surfaces of the substrate.  The saturation magnetization can be extracted by subtracting out the diamagnetic contribution.  Fig. \ref{polishing}(a) summarizes the whole set of measurements of the saturation moment versus unpolished surface area. 

 Energy Dispersive X-ray Spectroscopy (EDX) and X-ray Diffraction Spectra (XRD) were measured on both the unpolished and polished surfaces of the STO substrates. These measurements revealed no significant difference between the polished and unpolished surfaces except that the XRD lines were considerably wider for the stressed and disordered unpolished side of the substrates, as expected\cite{cullity}.     Data for  the unpolished surfaces are shown in Fig. \ref{EDX_XRD}  which reveal no impurity phases or magnetic elements to the level of sensitivity of these techniques.  The peak near 3.7 keV indicates the presence of a small amount of calcium in these substrates.

\subsection{Defect Ferromagnetism}

\begin{figure}
 \includegraphics{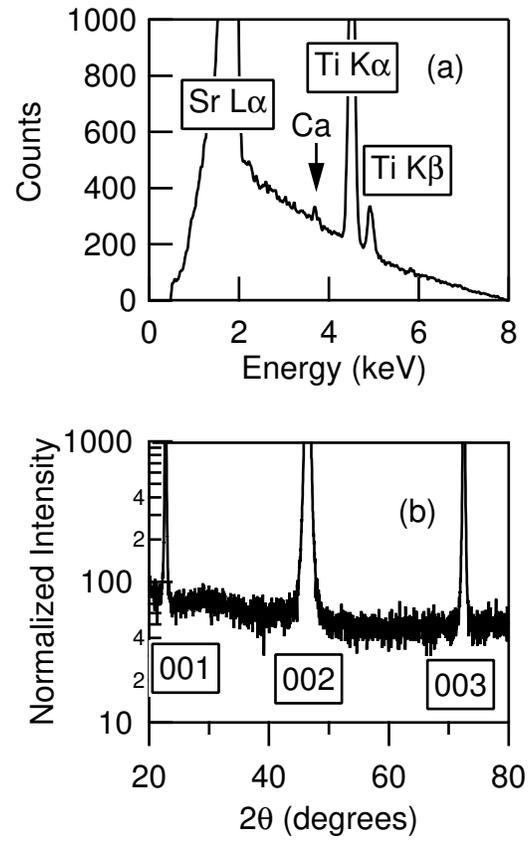}
\caption{Upper Panel (a): Energy Dispersive Xray Spectrum of the unpolished side of STO (001) substrate.  Lower Panel (b): Xray Diffraction pattern of the unpolished side of STO (001) substrate}
\label{EDX_XRD}
\end{figure}

There are at least  two possibilities for the FM hysteresis of the unpolished surfaces.  Firstly,  it could be due to particles of FM material left from either the diamond saw or wire saw cuts used to produce substrates from the crystal boule, or from handling by the suppliers of the unpolished edges using stainless steel tweezers.  At low enough concentration, these particles would not  be detected by the EDX and XRD measurements illustrated in Fig. \ref{EDX_XRD}.   Secondly, the hysteresis  could be  associated  with extended or point defects such as oxygen vacancies in the vicinity of the cut surface.  Two different approaches to this question were taken:  (i)  A systematic study was undertaken of the effect of annealing atmosphere and temperature on both the room temperature magnetic hysteresis  as well as the width of the XRD lines.  The reasoning was that annealing should heal some of the disorder simultaneously reducing the width of the XRD lines and remanent moment.  In addition, annealing in vacuum allows one to introduce a controlled density of oxygen vacancies via the annealing temperature.  (ii) Artificial disorder was introduced by Ar$^+$ or Ar$^{2+}$ bombardment of polished STO surfaces to see if a magnetic moment could be induced which would clearly be associated with disorder.  The goal was to find a way of introducing controlled amounts of disorder associated with surface ferromagnetism.

\begin{figure}
 \includegraphics{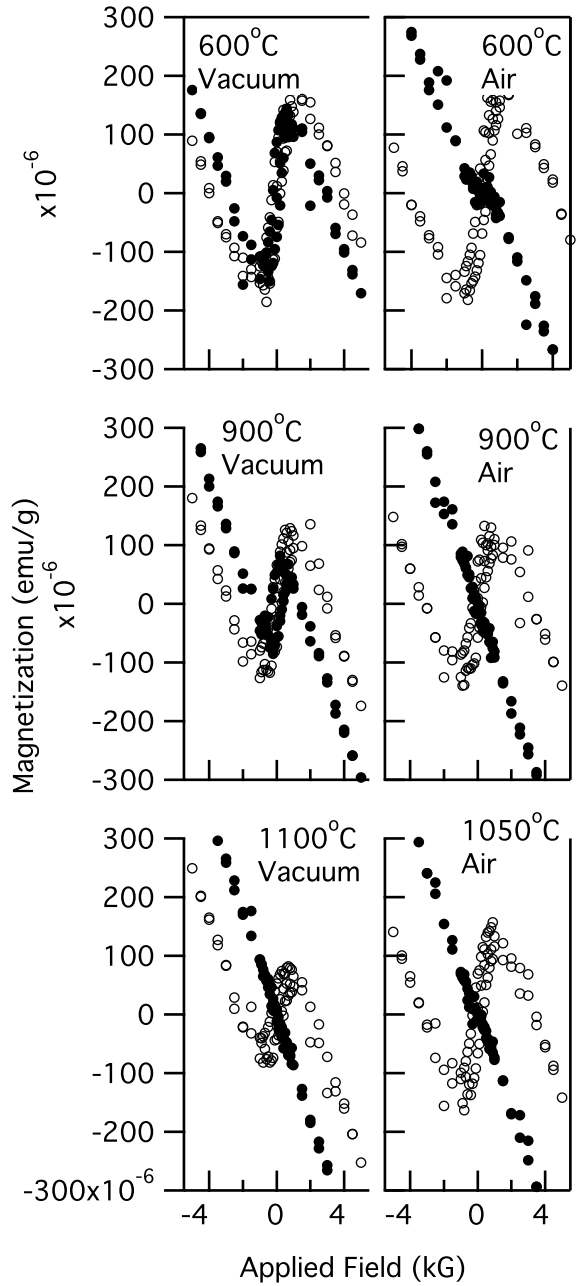}
\caption{Magnetization versus applied field at 295 K for  1sp STO substrates before (open circles) and after annealing (closed circles) at various temperatures and in different atmospheres.  Vacuum implies a pressure of approximately $5 \times 10^{-6}$ Torr.  }
\label{treatments}
\end{figure}

Annealing of the STO substrates was performed at different temperatures in either a vacuum of approximately $5 \times 10^{-6}$ Torr (reducing environment) or in air (oxidizing environment)  by placing the substrate in an Al$_2$O$_3$ crucible  inside a quartz evacuation chamber in the furnace.  A pattern emerged which is illustrated in  Fig. \ref{treatments}.   Annealing in air,  for $T > 600^\circ$C,  destroyed or significantly decreased the room temperature remanent magnetization while annealing in vacuum for $T<900^\circ$C did not destroy the remanent magnetization. 

\begin{figure}
 \includegraphics{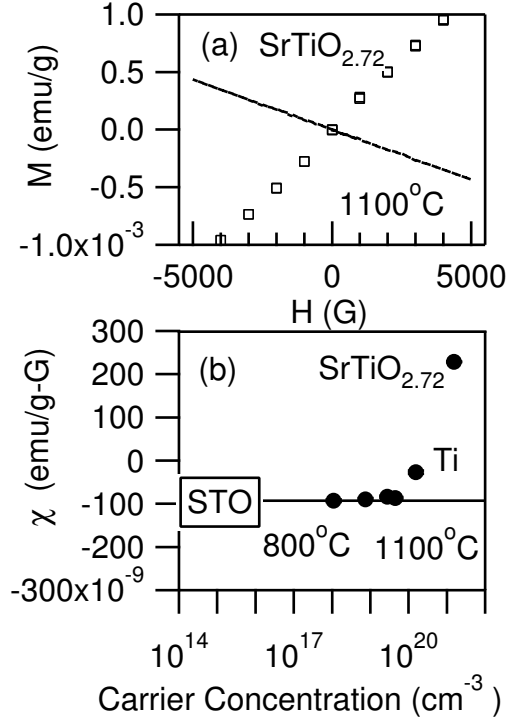}
\caption{(a)  Magnetization (M) versus applied field (H) at 295 K for a sample of SrTiO$_{2.72}$ and for a STO substrate that has been annealed at 1100C for 10 h in a vacuum of $5\times10^{-6}$ Torr.  The M vs. H curves of substrates with no unpolished surfaces annealed between 800 and 1100 $^\circ$C in vacuum are linear, but only one representative curve (1100$^\circ$C) is shown for clarity.  (b) Room temperature magnetic susceptibility of SrTiO$_{3-x}$ as a function of free carrier concentration - measured via the room temperature dc resistivity as discussed in Ref. \onlinecite{crandles} - which in a first approximation is proportional to oxygen vacancy density.  Points are included for a series of STO substrates annealed for 10 h in vacuum at 800,900,1000, and 1100C.   The point marked `Ti' is for a substrate annealed in a vacuum sealed quartz tube containing powdered Ti.  The point marked SrTiO$_{2.72}$ is for a grossly non-stoichiometric single crystal. The solid horizontal line indicates the measured room temperature susceptibility of pure SrTiO$_3$.}
\label{vary_reduction}
\end{figure}

The vacuum annealing experiments produced samples with a wide range of oxygen vacancy density.  Increasing the annealing temperature from 800$^\circ$C to 1100$^\circ$C  systematically decreased the dc resistivity (measured using the van der Pauw technique) and increased the dark colouration of the STO substrates. It is possible to produce darker, even less resistive samples by either annealing the substrate along with powdered Ti in a sealed, evacuated tube or by intentionally growing grossly oxygen deficient crystals \cite{gong}.  Sample magnetization data for one of the reduced substrates as well as that of a SrTiO$_{2.72}$ crystal are shown in Fig. \ref{vary_reduction} (a).   Note that data of  Fig. \ref{vary_reduction} (a) are for samples that have no unpolished surfaces.  The M vs. H relations of all the oxygen deficient samples were linear, while only two such curves are shown in   Fig. \ref{vary_reduction} (a) for simplicity.  This demonstrates clearly that oxgyen vacancies do not produce FM in STO. The vacancy densities were not directly measured but there is a well known correlation between the room temperature dc resisitivity and free carrier concentration\cite{crandles} where, in the simplest picture, each oxygen vacancy donates two free electrons to the conduction band.    In Fig. \ref{vary_reduction}(b),  the susceptibility of pure STO, which is due to the diamagnetic response of the closed shell ions plus a Van Vleck term\cite{candela},  is  $\chi=-9.2\pm0.5\times10^{-8}$ emu/g-G  and becomes less negative with the positive Pauli contribution of an increasing number of free carriers. This was the explanation given by Frederikse and Candela\cite{candela} and the data  illustrated in Fig. \ref{vary_reduction}(b)  agrees to within 10\%  with their earlier measurements\cite{candela}. Almost all of the samples included in Fig. \ref{vary_reduction}(b) have carrier concentrations compatible with low temperature superconductivity \cite{koonce}.  Because the samples which have carrier density ($\approx 10^{19}-10^{21}$cm$^{-3}$) compatible with superconductivity exhibit linear magnetic susceptibilities, it appears that STO is not a system where FM and superconductivity coexist. 

\begin{figure}
 \includegraphics{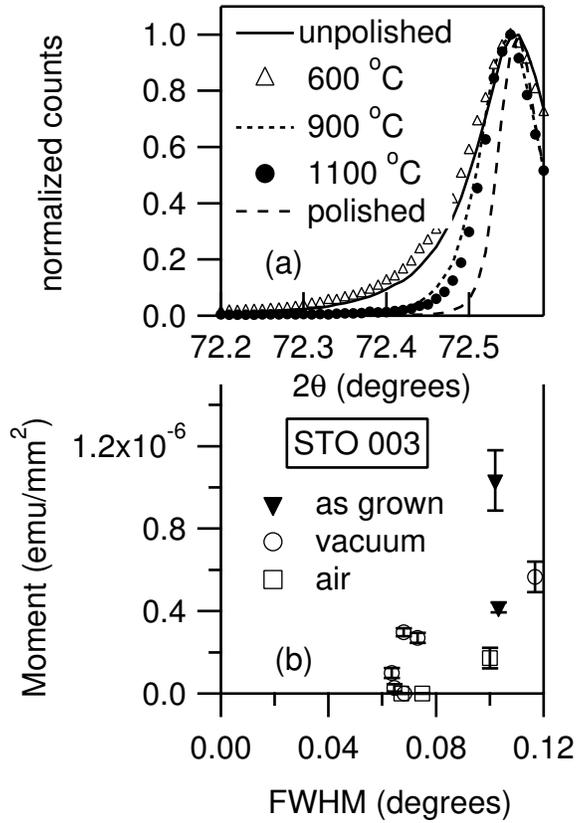}
\caption{Upper Panel (a): (003) line of STO XRD spectrum for substrates annealed in vacuum of 5$\times10^{-6}$ Torr at various temperatures or as purchased.  Except for the long dash curve which is for a polished side, all the spectra are for unpolished surfaces. The data are similar for air-annealed samples.  Lower Panel (b): Remanent moment at 295K versus the full width at half maximum of the (003) line of the XRD spectrum taken on unpolished surfaces  unnannealed and annealed (air or vacuum) at a variety of temperatures.  The uncertainty in moment is for a single measurement of the hysteresis curve on a single sample. }
\label{linewidth}
\end{figure}

Having concluded that oxgyen vacancies do not produce the FM hysteresis, one might hypothesize that other kinds of disorder such as dislocations and interstitials produced by the diamond saw cuts  may be associated with the FM hysteresis.  X-ray diffraction reveals that annealing  does indeed relieve the strain on the unpolished surfaces as shown in Fig. \ref{linewidth}(a) which illustrates how the linewidth of the (003) diffraction peak from the unpolished side decreases with annealing temperature.  The line on the polished side is also illustrated, for comparison.  The decrease in linewidth with annealing temperature is seen in all diffraction peaks for both vacuum and air annealing.      However,  Fig. \ref{linewidth}(b)  which is a plot of of the  remanent moment versus linewidth of the x-ray diffraction peaks,  suggests that the correlation between remanent moment and surface strain and disorder is not strong  not only because of the scatter in the data but because air annealing and vacuum annealing have such different effects on magnetization but produce similar decreases in the linewidth of the x-ray diffraction peaks

\begin{figure}
 \includegraphics{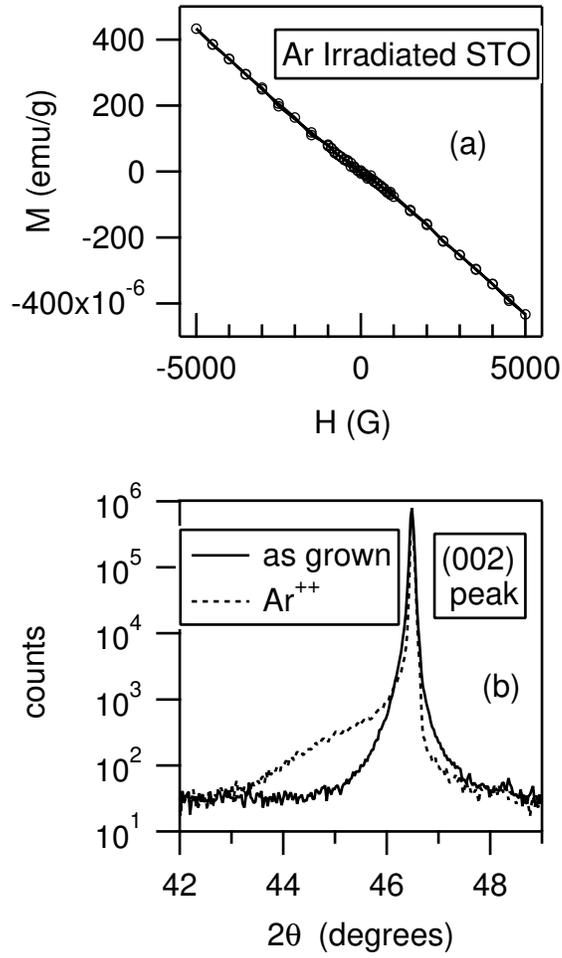}
\caption{(a) Linear magnetization versus applied field at 295 K for  an STO substrate that had been bombarded by a dose of 10$^{16}$  90 keV Ar$^{2+}$ ions.
(b) Portion of $2\theta$ xray scan comparing the diffraction peak before and after irradiation by 90 keV Ar$^{2+}$ ions}
\label{argon}
\end{figure}

As mentioned earlier, Ar ion bombardment was utilized to produce surface disorder in an attempt to mimic the disorder observed in the x-ray spectra of the unpolished surfaces.  This was done by taking samples with polished surfaces exhibiting linear magnetization versus applied field relations and mounting them with non-ferromagnetic adhesive onto 5-cm diameter Si wafers and placing them in an Ar ion beam to receive a dose of 10$^{16}$ ions.  Two beam energies were attempted: either 10 keV  (range 92\AA \ straggle 42\AA \ estimated using SRIM software (Stopping Range of Ions in Matter\cite{srim}))  or 90 keV (range 570\AA \ straggle 250\AA). A dose of 10$^{16}$ 90 keV ions produces a huge amount of damage, milling away part of the surface and producing an amorphous layer than can be observed using x-ray diffraction.  Fig. \ref{argon}(b) contrasts the (002) diffraction peaks from both irradiated and non-irradiated surfaces of the same substrate.  Note the creation of a broad peak at lower diffraction angle which is associated with the amorphous layer of thickness roughly 60 nm.  For comparison, diffraction peaks in amorphous Si can have widths well over one degree \cite{britton}.  Note that the main (002) peak of the Ar bombarded side of the crystal is narrower than that of the non-irradiated side.  On the non-irradiated but polished surface, x-rays are scattered from both a surface strained layer plus a less strained region deeper in the crystal.  On  the irradiated side, the narrow peak represents diffraction from deeper in the crystal than the surface amorphous layer.  More importantly, a lack of magnetic hysteresis was measured  after ion bombardment as shown in Fig. \ref{argon} (a) for both incident Ar ion kinetic energies.  It has been conclusively shown that FM is not associated with oxgyen vacancies, nor with an amorphous layer produced by Ar ion bombardment while Fig. \ref{linewidth}(b) offers only a weak correlation between surface strain and disorder and the observed saturation moment.

\begin{figure}
\includegraphics{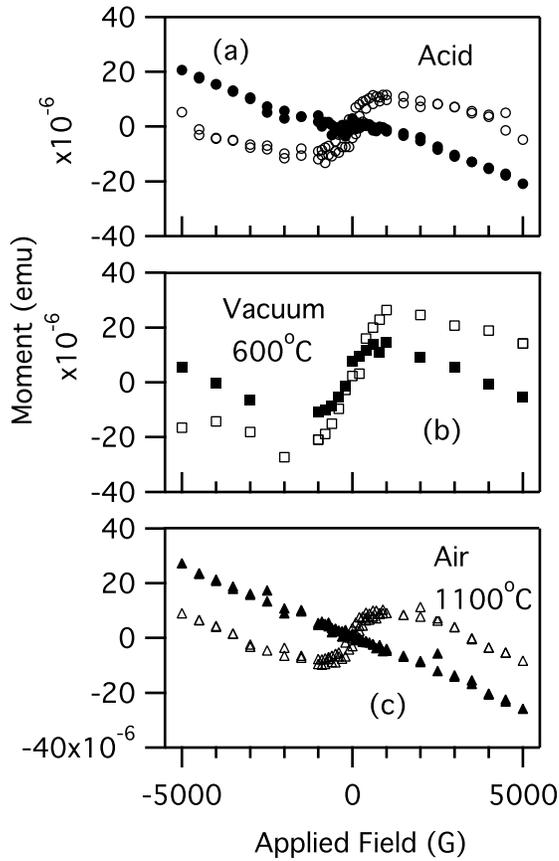}
\caption{All panels show  moment versus applied field at 295 K for $5\times5\times0.5$ mm STO substrates. The unfilled symbols show data  before the acid or annealing treatment while the filled symbols show the data after the treatment.  Panel (a):  Effect of Nitric Acid Treatment on as purchased  substrates.     Panel (b):  Effect of annealing at 600 $^\circ$C in a vacuum of $5 \times 10^{-6}$\ Torr on  STO substrate artificially  spiked with Fe.  Panel (c): Effect of high temperature air annealing (1100 $^\circ$C)  on  STO artificially spiked with Fe. }
\label{spiked}
\end{figure}

It is difficult to prove the second hypothesis,  in which the FM hysteresis of the unpolished surfaces is due to contamination by FM particles, because of the low  particle density. Recall that neither XRD nor EDX spectra illustrated in Fig. \ref{EDX_XRD} showed any evidence of impurity elements or phases.  The approach taken here was to see whether artificially contaminated samples produced the same annealing behaviour as the raw substrates. For the final set of experiments performed  samples  were prepared with no moment by soaking a 1sp substrate in nitric acid for 10 minutes in an ultrasonic bath as shown in Fig. \ref{spiked}(a).  Subsequently such samples were artificially contaminated with iron by rubbing the unpolished surface on stainless steel.  Comparison of Figs. \ref{spiked}(a),(b) and (c) demonstrates that artificial contamination creates a moment of roughly the same magnitude as observed in the as purchased 1sp substrates.  Finally the iron contaminated samples were subject to the same annealing treatments as untreated 1sp substrates.  Comparison of Fig. \ref{treatments} and Fig. \ref{spiked}(b) and (c) illustrates that the iron contaminated samples demonstrate the same annealing behaviour as the untreated 1sp substrates.

A natural explanation of Fig. \ref{treatments} and Fig. \ref{spiked} comes from a comparison of the magnetic properties of iron and the iron oxides which are listed in Table \ref{iron_oxides}.  Iron particles on the rough surfaces of the substrates coming from either diamond saw cuts or handling with stainless steel tweezers, are the source of the FM hysteresis observed on the unpolished surfaces. Below 900$^\circ$C, vacuum annealing would either maintain these particles in the Fe state,  or at 1100$^\circ$C there may oxidation to antiferromagnetic wustite FeO whose presence is indicative of a highly reducing environment \cite{chen}.   On the other hand air annealing would oxidize the iron particles to antiferromagnetic $\alpha-$Fe$_2$O$_3$ at all temperatures.    $\alpha-$Fe$_2$O$_3$ is more thermodynamically stable than $\gamma-$Fe$_2$O$_3$ \cite{cannas}. Definite confirmation of the presence of iron on unpolished surfaces awaits detection by more sensitive surface elemental probes like PIXE (particle induced x-ray emission).  It is of interest to compare the present work with the only other systematic study of the ferromagnetic properties of commercial substrates by Salzer {\it et al.} who considered sapphire (Al$_2$O$_3$) substrates purchased from Crystec\cite{salzer}.  The authors studied the magnetic hysteresis curves of both 1sp and 2sp samples of a variety of orientations and sizes.  Significantly they also tried to measure surface magnetic impurity concentration using PIXE (proton irradiated x-ray emission) measurements.  The PIXE measurements revealed that Fe was the main impurity  and that the average Fe concentration on unpolished c-faces was 230 ng/cm$^2$ while on polished c-faces the concentration was 10 ng/cm$^2$.  It is very interesting to note that a linear fit of Fig. \ref{polishing}(a) yields  Fe density of 190$\pm40$ ng/cm$^2$ if one assumes that the moment of the substrates is produced by pure Fe with a room temperature magnetization of 220 emu/g.

\section{Discussion}

There has been an on-going discussion surrounding ferromagnetism in magnetic ion free oxides. On one side are those who argue that ferromagnetism is produced by defects in the thin flims\cite{yoon_jcm,yoon,hong_tio2,rumaiz,sudakar, coey_prb,hong_zno}, while on the other side are those who argue that it is a spurious effect produced by iron contamination\cite{abraham,golmar, mofor}. Since some of the strongest claims linking oxygen vacancies and ferromagnetism have been made for   TiO$_{2-\delta}$ thin films, it is useful to compare the TiO$_{2-\delta}$  data with the present measurements on single crystal SrTiO$_3$.

A summary of the properties of TiO$_{2-\delta}$ thin films grown by a number of techniques on a variety of substrates appears in Table \ref{TiO2_films}.  The most notable feature of the data is that several groups have found that the saturation magnetization depends on the oxygen partial pressure during film deposition and have also observed that air annealing destroys the ferromagnetism.  All these authors have concluded that this is evidence that oxygen vacancies are producing the ferromagnetic hysteresis of the films and one has cited theoretical support\cite{han} for oxygen-vacancy-induced-ferromagnetism in TiO$_2$.   However, no measurements of oxygen vacancy were actually made in these works. 

On the other hand,  in the present work,  reasonable estimates of oxygen vacancy density can be inferred from the free carrier density  measurements made and Fig. \ref{vary_reduction} shows that  sample magnetization is a linear function of applied field even in samples with a  large density of bulk oxygen vacancies.  It seems very clear that oxygen vacancies do not produce ferromagnetism in STO.  The effect of vacuum annealing on  STO and TiO$_2$ is quite similar where oxygen vacancies produce an increase in the electrical conductivity.   Hence, the differences in ferromagnetism associated with oxygen vacancies in these two materials must be associated with differences in the Ti-O network where STO contains a network of corner shared  TiO$_6$ octahedra, while there is a combination of edge and corner sharing in rutile.

This is not the first work to have shown that commercial substrates without films themselves exhibit ferromagnetic hysteresis.   Saturation moments for various  substrates as measured by a number of groups are listed in Table \ref{commercial_substrates}.  It is interesting to note that aside from the anomalously large moment measured for LaAlO$_3$ by  Hong {\it et al.}\cite{hong_tio2},  there appears to be a rough correlation between substrate moment and the hardness of the substrate which are all harder than stainless steel.   It is difficult to compare the moments of the TiO$_{2-\delta}$ films with the substrates because usually it is the magnetization of the samples, rather than the raw moments that are listed in the papers.  However, it is curious that both Yoon {\it et al.}\cite{yoon_jcm,yoon} and Sudakar {\it et al.}  \cite{sudakar}  find that the magnetization to be a decreasing function of film thickness.   This data could be partially explained by iron contamination on the unpolished substrate surfaces.   It is also puzzling  that Sudakar  {\it et al.}  measure magnetization of films produced by reactive RF sputtering to be ten times less than films produced by pulsed laser deposition.  They suggest that it is because the polycrystalline sputtered films on fused quartz may have less defects than the epitaxial PLD films on LaAlO$_3$, but this does not seem convincing.  A simpler explanation could be that different substrates have different amounts of surface contamination.    It is also interesting that Yoon {\it et al.}, found the Curie temperature of TiO$_{2-\delta}$ thin film samples  to be 880 K which is close to the transition temperature of $\gamma-$Fe$_2$O$_3$ as shown in Table \ref{iron_oxides}.   The present annealing experiments illustrated in Fig. \ref{treatments},  which can be explained by the presence of iron on the unpolished surfaces of substrates,  raise the possibility  that the supposed dependence of saturation moment on oxygen partial pressure in the TiO$_2$ films might be the response of  the substrates themselves.  

\section{Conclusions}

Ferromagnetic hysteresis with a saturation moment of the order of $5\times 10^{-5}$ emu/cm$^2$ is regularly observed in STO (001) substrates with 1 or 2 sides polished as purchased from a variety of suppliers.   Oxygen vacancies do not appear to be associated with ferromagnetism in STO nor can a magnetic moment be created by Ar ion bombardment.  Reduced STO is not a system where FM and superconductivity coexist. In the as purchased samples, the hysteresis is associated with the unpolished surfaces and is likely due to Fe contamination  from handling with stainless steel tweezers or the diamond or wire saw cuts rather than oxygen vacancies or other lattice defects.  It is possible that all single crystal substrates used for thin film growth  contain ferromagnetic iron or iron oxide on the unpolished surfaces.   Polishing and simple acid washes can remove the weak ferromagnetic signal.  Extreme care must be taken in handling samples to avoid contamination and to interpret magnetic data measured on thin films grown on STO substrates with unpolished surfaces. One possible application of the results of this paper is for thin film growers using STO substrates.  By annealing at high temperature before film growth, all traces of the ferromagnetic signal of the substrate can be removed giving film growers confidence that any subsequent ferromagnetic signals are due to the films themselves.   It may also be of interest to further study iron implantation as a way to produce ferromagnetic surface layers in a controlled manner.

\newpage

%
%

%
%
\begin{table}
\caption{Properties of iron and iron oxides. $^a$ FeO is typically formed in reducing atmospheres.  $^b$The magnetization of $\alpha$-Fe$_2$O$_3$ does not saturate. The value listed is for T=295 K and H=1T , well above any fields used in this study. }
\ \\
\begin{tabular}{|c|c|c|c|c|}
\hline
Material 				& Magnetic Order 	& T$_c$ or T$_N$\ (K) & M$_{sat}$(300K) emu/g	& References\\ 
\hline
Fe 			&  Ferromagnetic	& 1043  & 	220 &\onlinecite{kittel}\\ 
FeO			& Antiferromagnetic & 200 &$^ a$ & \onlinecite{chen}\\
Fe$_3$O$_4$             & Ferrimagnetic         & 858  	&	30-90 & \onlinecite{kittel,goya,zaitsev,strobel}\\
$\gamma$-Fe$_2$O$_3$ 	& Ferrimagnetic		& 878 	&	50-75 	& \onlinecite{strobel}\\
$\alpha$-Fe$_2$O$_3$ 	& Antiferromagnetic	& 955 	&	0.5$^b$ 	& \onlinecite{bodker,lin}\\

\hline
\end{tabular}
\label{iron_oxides}
\end{table}


%
%
\begin{table}
\caption{Summary of data collected on TiO$_{2-\delta}$ thin films. PLD  signifies pulsed laser deposition.  MOD signifies metal-organic decomposition.  RRFS signifies reactive radio frequency sputtering. \\  $^a$Annealing in O$2$ for 8h at 650 C destroyed the FM. $^b$The substrates were held at 27$^\circ$ C for RF sputtering in an oxygen partial pressure of 1.5 mTorr using Ar as the sputtering gas. These films were then annealed at 900C in vacuum to obtain rutile structure.  }
\ \\
\begin{tabular}{|c|c|c|c|c|c|c|}
\hline
Method 	& Substrate & Deposition T (C)  & P(O$_2$) & Thickness (nm) & M$_{sat}$(300K)  & References\\ 
 	& &  (C) &  (mTorr) & (nm)&  emu/cm$^3$ & \\ 
\hline
PLD & LaAlO$_3$ & 700 & 0.3 & 250 & 33 & \onlinecite{yoon_jcm,yoon}\\
 &  &  & 25 & 250 &  12 & \\
\hline
PLD 	& SiO$_2$ 	& 800 	& 0.02 	& na & 0.32 & \onlinecite{rumaiz}\\
 	&  		&  	& 0.2 	& na & 0.03 & \\
\hline
PLD & LaAlO$_3$ & 600-700 & 0.001 & 200 & 20 & \onlinecite{hong_tio2}\\
   &           &          & 0.001 & 200 & 0$^a$ & \\
\hline

RRFS & fused quartz & 27$^b$ & 1.5 & 25 &  40 & \onlinecite{sudakar}\\
& & & &  50 &  10 &\\
  &  &  &  & 100 &  5 & \\

\hline
MOD & Al$_2$O$_3$ & 500  & air & 500-900 &  0.8 & \onlinecite{sudakar}\\
 & 2sp -  25 mm$^2$ & &  &  &   & \\

\hline

\hline
\end{tabular}
\label{TiO2_films}
\end{table}

%
%
\begin{table}
\caption{Magnetic moments measured in  5mm$\times$5mm$\times$0.5 mm (1sp) commercial substrates.  YSZ is yttria stabilized zirconia .  It is of interest that the moh's hardness of  stainless steel is listed between 5 and 6. }
\ \\
\begin{tabular}{|c|c|c|c|}
\hline
Material 				& 300 K Saturation Moment (emu) & Moh's Hardness & References\\ 
\hline
MgO & $1\times10^{-6}$ & 5.8 & Ref. \onlinecite{hong_tio2}\\
LaAlO$_3$ & $1\times10^{-5}$ & 5-6 & Ref. \onlinecite{hong_tio2}\\
LaAlO$_3$ & $1\times10^{-6}$ & 5-6 & Ref. \onlinecite{golmar}\\
SrTiO$_3$ & $1.6\pm0.6\times10^{-5}$ & 6-6.5& this work \\
YSZ & $2.5\times10^{-5}$ & 9 & Ref. \onlinecite{hong_tio2} \\
Al$_2$O$_3$ & $3\pm1\times10^{-5}$ & 9 & Ref. \onlinecite{mofor} \\
Al$_2$O$_3$ & $3\pm2.5\times10^{-5}$ & 9 & Ref. \onlinecite{salzer} \\

\hline
\end{tabular}
\label{commercial_substrates}
\end{table}

%
%
\newpage


\end{document}